# Precision and accuracy of single-molecule FRET measurements – a worldwide benchmark study

Björn Hellenkamp[1,°], Sonja Schmid[1,°], Olga Doroshenko[20], Oleg Opanasyuk[20], Ralf Kühnemuth[20], Soheila Rezaei Adariani[15], Anders Barth[21], Victoria Birkedal[9], Mark E. Bowen[11], Hongtao Chen[26], Thorben Cordes[14,25], Tobias Eilert[19], Carel Fijen[7], Markus Götz[1], Giorgos Gouridis[14,25], Enrico Gratton[26], Taekjip Ha[22], Christian A. Hanke[20], Andreas Hartmann[17], Jelle Hendrix[5,6], Lasse L. Hildebrandt[9], Johannes Hohlbein[7,8], Christian G. Hübner[16], Eleni Kallis[19], Achillefs N. Kapanidis[10], Jae-Yeol Kim[23], Georg Krainer[17,18], Don C. Lamb[21], Nam Ki Lee[23], Edward A. Lemke[3], Brié Levesque[11], Marcia Levitus[24], James J. McCann[11], Nikolaus Naredi-Rainer[21], Daniel Nettels[4], Thuy Ngo[22], Ruoyi Qiu[12], Carlheinz Röcker[19], Hugo Sanabria[15], Michael Schlierf[17], Benjamin Schuler[4], Henning Seidel[16], Lisa Streit[19], Philip Tinnefeld[13,27], Swati Tyagi[3], Niels Vandenberk[5], Keith R. Weninger[12], Bettina Wünsch[13], Inna S. Yanez-Orozco[15], Jens Michaelis[19,*], Claus A.M. Seidel[20,*], Timothy D. Craggs[2,10,*], Thorsten Hugel[1,*]

[1]Institute of Physical Chemistry, University of Freiburg, Germany

[2]Department of Chemistry, University of Sheffield, S3 7HF, UK

[3]Structural and Computational Biology Unit, EMBL Heidelberg, Germany

[4]Department of Biochemistry, University of Zurich, Switzerland

[5]Laboratory for Photochemistry and Spectroscopy, Department of Chemistry, University of Leuven, B-3001 Leuven, Belgium

[6]Present address: Hasselt University, Faculty of Medicine and Life Sciences and Biomedical Research Institute, B-3590 Hasselt, Belgium

[7]Laboratory of Biophysics, Wageningen University & Research, 6708 WE, Wageningen, NL

[8]Microspectroscopy Research Facility Wageningen, Wageningen University & Research, 6708 WE, Wageningen, NL

[9]Interdisciplinary Nanoscience Center and Department of Chemistry, Aarhus University, 8000 Aarhus C, Denmark

[10]Gene Machines group, Clarendon Laboratory, Department of Physics, University of Oxford, Parks Road, OX1 3PU, Oxford, UK

[11]Department of Physiology & Biophysics, Stony Brook University, Stony Brook, NY 11794, USA

[12] Department of Physics, North Carolina State University, Raleigh, NC 27695 USA

[13]Institute of Physical & Theoretical Chemistry, and Braunschweig Integrated Centre of Systems Biology (BRICS), and Laboratory for Emerging Nanometrology (LENA), Braunschweig University of Technology, Rebenring 56, 38106 Braunschweig, Germany

[14]Molecular Microscopy Research Group, Zernike Institute for Advanced Materials, University of Groningen, Nijenborgh 4, 9747 AG Groningen, The Netherlands

[15]Department of Physics and Astronomy, Clemson University, 29634 Clemson, SC. USA

[16]Institute of Physics, University of Lübeck, Germany

[17]B CUBE – Center for Molecular Bioengineering, TU Dresden, Arnoldstr. 18, 01307 Dresden, Germany

[18]Molecular Biophysics, University of Kaiserslautern, Erwin-Schrödinger-Str. 13, 67663 Kaiserslautern, Germany.

[19]Institute for Biophysics, Ulm University, Albert-Einstein-Allee 11, 89081 Ulm, Germany

[20] Chair for Molecular Physical Chemistry, Heinrich-Heine-Universität Düsseldorf, 40225 Düsseldorf, Germany

[21]Lamb lab, Department of Chemistry, LMU München, Butenandtstrasse 5-13, 81377 Germany

[22]Ha lab, Department of Biomedical Engineering, John Hopkins University, USA

[23]School of Chemistry, Seoul National University, Seoul, South Korea

[24]School of Molecular Sciences and The Biodesign Institute, Arizona State University, USA

[25]Physical and Synthetic Biology, Faculty of Biology, Ludwig Maximilians-Universität München, Großhadernerstr. 2-4, 82152 Planegg-Martinsried, Germany

[26]Department of Biomedical Engineering, University of California, Irvine, Irvine, California, USA

[27]Department of Chemistry, Ludwig-Maximilians-Universitaet Muenchen, Butenandtstr. 5-13, 81377 München, Germany.

° these authors contributed equally

*corresponding authors: jens.michaelis@uni-ulm.de; cseidel@hhu.de; t.craggs@sheffield.ac.uk; thorsten.hugel@pc.uni-freiburg.de

**Abstract**

Single-molecule Förster resonance energy transfer (smFRET) is increasingly being used to determine distances, structures, and dynamics of biomolecules *in vitro* and *in vivo*. However, generalized protocols and FRET standards ensuring both the reproducibility and accuracy of measuring FRET efficiencies are currently lacking.

Here we report the results of a worldwide, comparative, blind study, in which 20 labs determined the FRET efficiencies of several dye-labeled DNA duplexes. Using a unified and straightforward method, we show that FRET efficiencies can be obtained with a standard deviation between ΔE = ±0.02 and ±0.05. We further suggest an experimental and computational procedure for converting FRET efficiencies into accurate distances. We discuss potential uncertainties in the experiment and the modelling. Our extensive quantitative assessment of intensity-based smFRET measurements and correction procedures serve as an essential step towards validation of distance networks with the ultimate aim to archive reliable structural models of biomolecular systems obtained by smFRET-based hybrid methods.

**Introduction**

Förster Resonance Energy Transfer (FRET) [1], also sometimes termed Fluorescence Resonance Energy Transfer, has become a well-established method for studying biomolecular conformations and dynamics at both the ensemble [2-4] and the single-molecule (sm) level [5-10]. In such experiments, the energy transfer between a donor and an acceptor fluorophore pair is quantified with respect to their proximity [1]. The fluorophores are usually attached via flexible linkers to defined positions of the system under investigation. Amongst other factors, the transfer efficiency depends on the inter-dye distance, which is well described by Förster's theory for distances > 30 Å [11,12]. Accordingly, FRET has been termed a 'spectroscopic ruler' on the molecular scale[2]. Such a ruler is an important tool to determine distances *in vitro,* and even in cells [13], with potentially Ångström accuracy and precision. In its single-molecule implementation, FRET largely overcomes ensemble- and time-averaging and can uncover individual species within heterogeneous and dynamic biomolecular complexes, as well as transient intermediates[5].

The two most popular smFRET approaches to determine distances are confocal microscopy on freely diffusing molecules, and total internal reflection fluorescence (TIRF) microscopy on surface-attached molecules. Here, we focus on intensity-based measurements, in which the FRET efficiency, *E,* is determined from donor and acceptor photon counts, and then subsequently used to calculate the inter-fluorophore distance according to Förster's theory.

The vast majority of intensity-based smFRET studies to date rely on characterizing relative changes in FRET efficiency. This ratiometric approach is often sufficient to distinguish different conformations of a biomolecule (e.g. an open conformation with low FRET efficiency vs. a closed conformation with high FRET efficiency), and to determine their interconversion kinetics. Yet, determining distances provides additional information that can be used, for example, to compare with known structures, or assign conformations to different structural states. In combination with prior structural knowledge and computer simulations, FRET-derived distances are increasingly being used to generate novel biomolecular structural models using hybrid structural tools [7-9,14-17].

However, comparing and validating distance measurements from different labs is difficult, especially given the lack of detailed methodological descriptions in many publications. In addition, different methods for data acquisition and analysis, often using home-built microscopes with in-house software, can have very different uncertainties and specific pitfalls. To overcome these issues, we have developed general methodological recommendations and well-characterized FRET-standard samples to enable the validation of results and the estimation of distance accuracy and precision. This approach should allow the scientific community to confirm consistency of smFRET-derived distances and structural models. To facilitate data validation across the field, we recommend the reporting of specific FRET-related parameters with a unified nomenclature.

Various fluorescence intensity- and lifetime-based procedures have been proposed with the aim of determining FRET efficiencies [10,18-24]. Here, we present a general methodological guide based on some of these procedures, that allows us to obtain quantitative and reliable smFRET data. The step-by-step procedure results in a consistent determination of FRET efficiencies that corrects for dye and setup characteristics and is independent of the software used. It includes deconvolution of the underlying uncertainties involved in the determination of the necessary correction factors, enabling scientists to specify the overall uncertainty of the determined FRET efficiency. These steps are tested in a worldwide, comparative, blind study by 20 participating labs. For standardized FRET samples, we show that FRET efficiencies can be determined with a standard deviation $\Delta E < \pm 0.05$.

In order to convert the measured smFRET efficiency to a distance, the Förster equation is used (*Table 1*, eq. (III)), which critically depends on the dye-pair-specific Förster radius, $R_0$. We discuss the measurements required to determine $R_0$ and the associated uncertainties. Another uncertainty arises from the fact that many positions are being sampled by the dye relative to the biomolecule to which it is attached. Therefore, specific models are used to describe the dynamic movement of the dye molecule, during the recording of each FRET-efficiency measurement [15,16]. In summary, the investigation of the underlying error sources enables us to specify uncertainties for individual FRET-derived distances.

We anticipate that the investigated samples and the presented procedure will help unify the research field, serving as a standard for future publications and benchmarking the use of smFRET as an accurate spectroscopic ruler.

# Results

We have chosen double-stranded DNA as a FRET standard for the following reasons: any DNA sequence can be synthesized; FRET dyes can be specifically tethered at desired positions; the structure of B-form DNA is well characterized; and the samples are stable at room temperature for a time window that is large enough for shipping to labs around the world. The donor and acceptor dyes are attached via C2- or C6-amino-linkers to thymidines of opposite strands (see *Supplementary Figure 1*). These thymidines were separated either by 15 or by 23 base pairs (*Figure 1* and *Online Methods section 1*). The attachment positions were known only to the reference lab that designed the samples. Based on the resulting high-efficiency and low-efficiency samples we were able to determine all correction parameters and to perform a self-consistency test (see below).

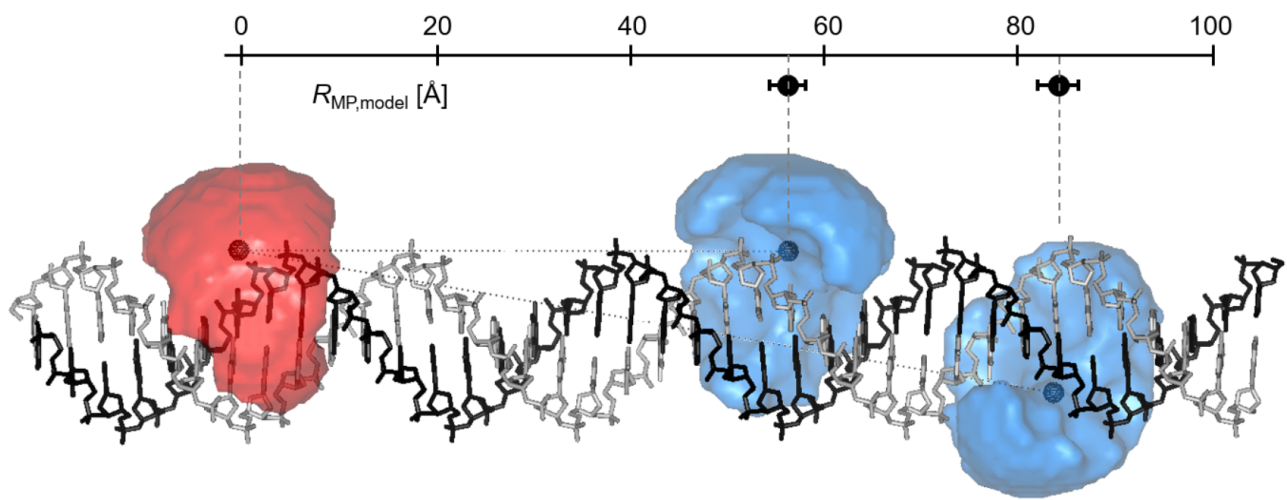

*Fig. 1:* Schematic of the FRET standard molecules. Double-stranded DNA is labeled with a FRET pair at 15 or 23 base-pair separation (for sequences see Online Methods). One DNA strand labeled with an acceptor dye (red) and the complementary strand labeled with a donor dye (blue) at one of two positions were annealed to generate the FRET standards. The accessible volumes (AVs) of the dyes are illustrated as semi-transparent surfaces and were calculated using freely available software [8]. The mean dye positions are indicated by darker spheres (assuming homogenously distributed dye positions, see Supplementary Note 2). The distance between the mean dye positions is defined as $R_{MP,model}$. Calculated values for $R_{MP,model}$ together with error bars obtained by varying parameters of the AV model are displayed (see Supplementary Note 2). The B-DNA model was generated using the Nucleic Acid Builder version 04/17/2017 for Amber [25].

A wide variety of dyes are used in smFRET studies. Here we used Alexa and Atto dyes (*Supplementary Figure 1*) due to their high quantum yields and well-studied characteristics. Eight hybridized double-stranded FRET samples were shipped to all participating labs. In the main text, we focus on four FRET samples that were measured by most labs in our study (see *Online Methods section 1* for details):

**1-lo**: Atto550/Atto647N, 23 bp-separation.

**1-hi**: Atto550/Atto647N, 15 bp-separation.

**2-lo**: Atto550/Alexa647, 23 bp-separation.

**2-hi**: Atto550/Alexa647, 15 bp-separation.

In this nomenclature, the number refers to the dye pair and the last two letters indicate either the low-efficiency (lo) or high-efficiency (hi) configurations. The results with other FRET pairs (Alexa488/Alexa594 and Alexa488/Atto647N) at these positions are reported in *Supplementary Figure 2* and *Supplementary Note 1*. As a first test for the suitability of the labels, we checked the fluorescence lifetimes and time-resolved anisotropies (*Supplementary Table 1*) of all donor-only and acceptor-only samples. The results indicate that there is no significant quenching and that all dyes are sufficiently mobile at these positions (see *Supplementary Note 1*).

There are two main routes to measuring accurate distances by FRET: fluorescence intensity-based and fluorescence lifetime-based measurements. In this study, we focus on the intensity-based methods, because they are easier to implement and were performed by more labs in our blind study. Specifically, we discuss solution-based measurements collected using a confocal microscope, and surface-based measurements collected using a TIRF microscope. In the framework of this study, other measurements were also performed using different fluorophores (samples 3 and 4) and different FRET methods (ensemble lifetime [26], single-molecule lifetime [20], and a phasor approach [9]; for these results, see *Supplementary Figure 2* and *Supplementary Note 1 and 5*).

Most intensity-based FRET measurements are performed on custom-built setups for single-molecule detection featuring at least two separate spectral detection channels for donor and acceptor emission (*Supplementary Figures. 3 and 4*). Here, the main challenge is the measurement of absolute corrected fluorescence intensities. The ideal solution is a ratiometric approach which for intensity-based confocal FRET measurements was pioneered by Weiss and coworkers using alternating two color laser excitation (ALEX) with microsecond pulses[21,27]. In this approach the total fluorescence signal (donor and acceptor emission) after donor excitation is normalized to the acceptor fluorescence after acceptor excitation, to correct for dye and instrumental properties [21]. The ALEX approach was also adapted for TIRF measurements [24]. To increase time resolution and to enable time-resolved spectroscopy, Lamb and coworkers introduced pulsed interleaved excitation (PIE) with picosecond pulses [28].

In both confocal and TIRF microscopy, the corrected FRET efficiency histogram is determined first, from which the expectation value of the FRET efficiency $\langle E \rangle$ is computed. Subsequently, the distance is calculated, assuming a suitable model for the inter-dye distance distribution and dynamics [6,11,29]. Below, we describe a concise and robust procedure that is suitable for both confocal and TIRF-based measurements. The results of our blind study underline the robustness and precision of this method. Further, we derive self-consistency arguments and comparisons to structural models and thereby confirm the accuracy of this method.

**<u>Procedure to determine the experimental FRET efficiency $\langle E \rangle$</u>**

Our general procedure is largely based on the Lee et al. approach [21], with modifications to establish a robust workflow and standardize the nomenclature. Intensity-based determination of FRET efficiencies requires the consideration of certain correction factors (see Table 1): Background signal correction (*BG*) from donor and acceptor channels; factor for spectral crosstalk ($\alpha$), arising from donor fluorescence leakage in the acceptor channel; factor for direct excitation ($\delta$) of the acceptor with the donor laser; detection correction factor ($\gamma$). The optimal way to determine these factors is by alternating the excitation between two colors, which allows for the determination of the FRET efficiency (*E*) and the relative stoichiometry (*S*) of donor and acceptor dyes, for each single-molecule event. This introduces the additional excitation correction factor ($\beta$) to normalize to equal excitation rates (see *Online Methods section 2.6*).

The following step-by-step guide for intensity-based FRET data analysis is subdivided for the type of FRET experiment (confocal and TIRF); notably, the order of the steps is crucial for the correct application of this procedure.

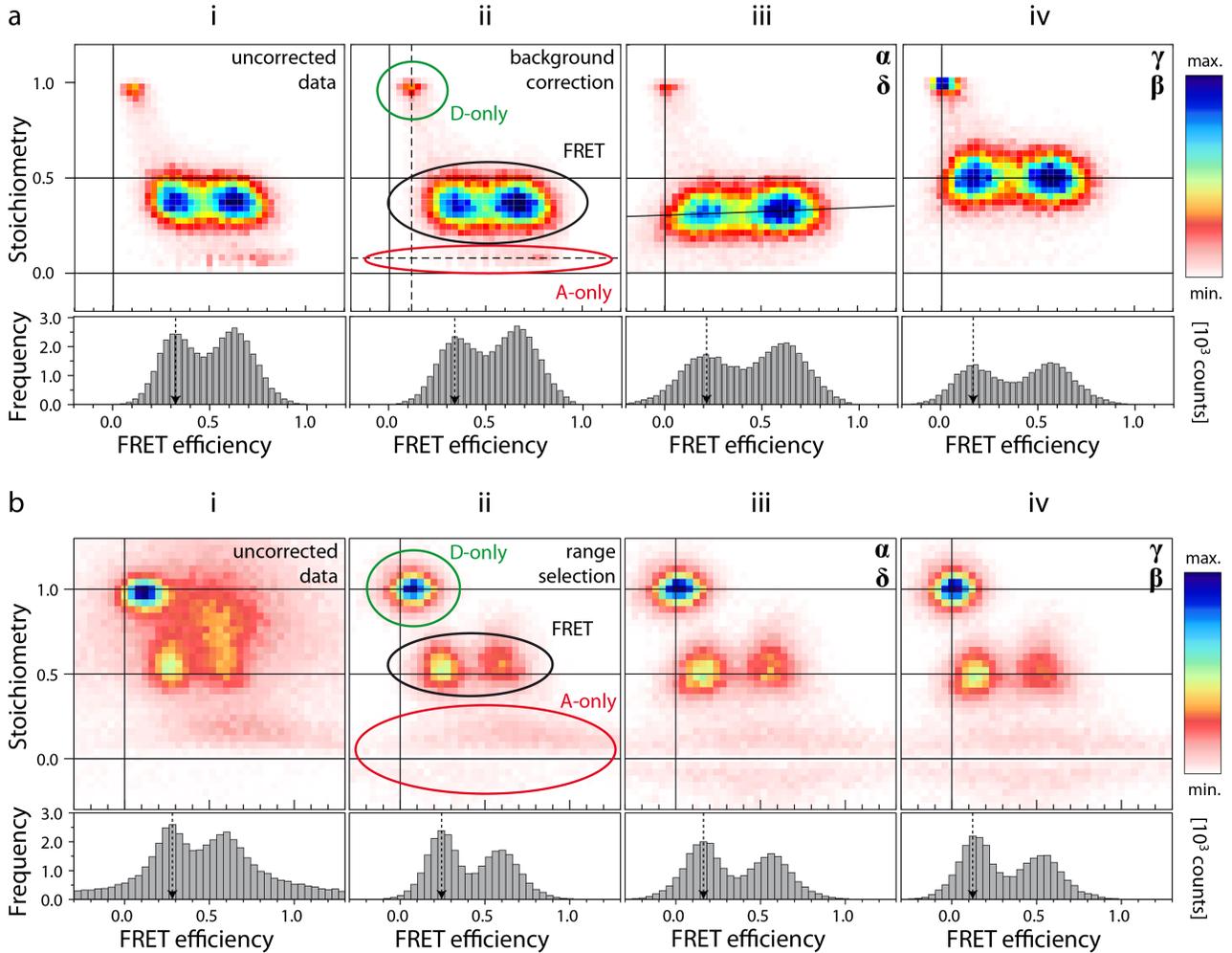

***Fig. 2:*** *Stepwise correction of confocal **(a)** or TIRF **(b)** data, shown for the combination of sample 1-lo and 1-hi. The general terms stoichiometry S and FRET efficiency E are used here in place of the corresponding specific terms for each correction step (see Online Methods – section 2). (i) $^{i}S_{app}$ vs $^{i}E_{app}$, (ii) $^{ii}S_{app}$ vs $^{ii}E_{app}$, (iii) $^{iii}S_{app}$ vs $^{iii}E_{app}$, (iv) S vs E. **(a)** Workflow for correcting the confocal data for background (i->ii), leakage (factor α) and direct excitation (δ) (ii->iii), excitation and detection factors (β,γ) (iii->iv). **(b)** Workflow for correcting the TIRF data for background and photo-bleaching by selection of the pre-bleached range (i->ii), leakage and direct excitation (ii->iii), detection and excitation factors (iii->iv). The efficiency histograms below show a projection of the data with a stoichiometry between 0.3 and 0.7. Note the significant shift of the FRET efficiency peak positions, especially for the low FRET efficiency peak (E~0.25 uncorrected to E~ 0.15 fully corrected). Donor only (D-only), FRET and acceptor only (A-only) populations are specified.*

*Diffusing molecules: Confocal Microscopy*

In confocal-based smFRET experiments with alternating excitation, photon bursts from individual molecules freely diffusing through the laser focus of a confocal microscope are collected and analyzed. From the data, first a 2D histogram of the uncorrected FRET efficiency ($^iE_{app}$) versus uncorrected stoichiometry ($^iS_{app}$) is calculated (*Figure 2a(i)*). See *Online Methods (section 2.0)* for the detailed procedure. Then, the average number of background photons is subtracted for each channel separately (*Figure 2a(ii)*). Next, to obtain the FRET sensitized acceptor signal ($F_{A|D}$), donor leakage ($\alpha\,^{ii}I_{Dem|Dex}$) and direct excitation ($\delta\,^{ii}I_{Aem|Aex}$) must be subtracted from the acceptor signal after donor excitation. As samples never contain 100% photoactive donor and acceptor dyes, the donor- and acceptor-only populations are selected from the measurement and used to determine the leakage and direct excitation (*Figure 2a(iii)*). After this correction step, the donor-only population should have an average FRET efficiency of 0 and the acceptor-only population should have an average stoichiometry of 0.

The last step deals with the detection correction factor $\gamma$ and the excitation correction factor $\beta$. If at least two species (two different samples or two populations within a sample) with different inter-dye distances are present, they can be used to obtain the "*global $\gamma$-correction*". If one species with significant distance fluctuations, e.g. through intrinsic conformational changes, is present a "*single-species $\gamma$-correction*" may be possible. Both correction schemes assume that the fluorescence quantum yields and extinction coefficients of the dyes are independent of the attachment point (see *Online Methods section 2.6*). The correction factors obtained by the reference lab are compiled in Table 2. The final corrected FRET efficiency histograms are shown in *Figure 2a(iv)*. The expected efficiencies $\langle E \rangle$ are obtained as the mean of a Gaussian fit to the respective efficiency distributions (see *Online Methods section 2.7*).

*Surface-attached molecules: TIRF Microscopy*

The correction procedure for TIRF-based smFRET experiments is similar to the procedure for confocal-based experiments. In the procedure used for ALEX data [24], a 2D histogram of the uncorrected FRET efficiency ($^iE_{app}$) versus uncorrected stoichiometry ($^iS_{app}$) is first generated (*Figure 2b(i)*). The background subtraction is critical in TIRF microscopy as it can contribute significantly to the measured signal. Different approaches can be used to accurately determine the background signal (see *Online Methods section 2.3*), such as measuring the background in the vicinity of the selected particle or measuring the intensity after photobleaching (*Figure 2b(ii)*). After background correction, the leakage and direct excitation can be calculated from the ALEX data as for confocal microscopy (*Figure 2b(iii)*).

Again, determination of the correction factors $\beta$ and $\gamma$ are critical[19]. As in confocal microscopy, one can use the stoichiometry information available from ALEX when multiple populations are present to determine an average detection correction factor (*global $\gamma$-correction*). In TIRF microscopy, the detection correction factor can also be determined on a molecule-by-molecule basis, provided the acceptor photobleaches before the donor (*individual $\gamma$-correction*). In this case, the increase in the fluorescence of the donor can be directly compared to the intensity of the acceptor before photobleaching. A 2D histogram of the corrected FRET efficiency versus the corrected stoichiometry is shown in *Figure 2b(iv)*.

In the absence of alternating excitation, the following problems were occasionally encountered during this study: (i) the low-FRET efficiency values were shifted systematically to higher efficiencies, because FRET efficiency values at the lower edge are overlooked due to noise; (ii) the direct excitation was difficult to detect and correct for, due to its small signal to noise ratio; (iii) the acceptor bleaching was difficult to detect for low FRET efficiencies. Therefore, implementation of ALEX is strongly recommended for obtaining accurate FRET data.

Nine of the twenty participating labs determined FRET efficiencies by confocal methods for sample 1 and 2 (*Figure 3a*). Seven of the twenty participating labs determined FRET efficiencies by TIRF-based methods and these are summarized in *Figure 3b*. The combined data from all labs measuring samples 1 and 2 agree very well, with a standard deviation for the complete data set of $\Delta E<\pm0.05$. This is a remarkable result, considering that different setup types were used (confocal- and TIRF-based setups) and different correction procedures were applied (e.g. individual, global or single species $\gamma$-correction).

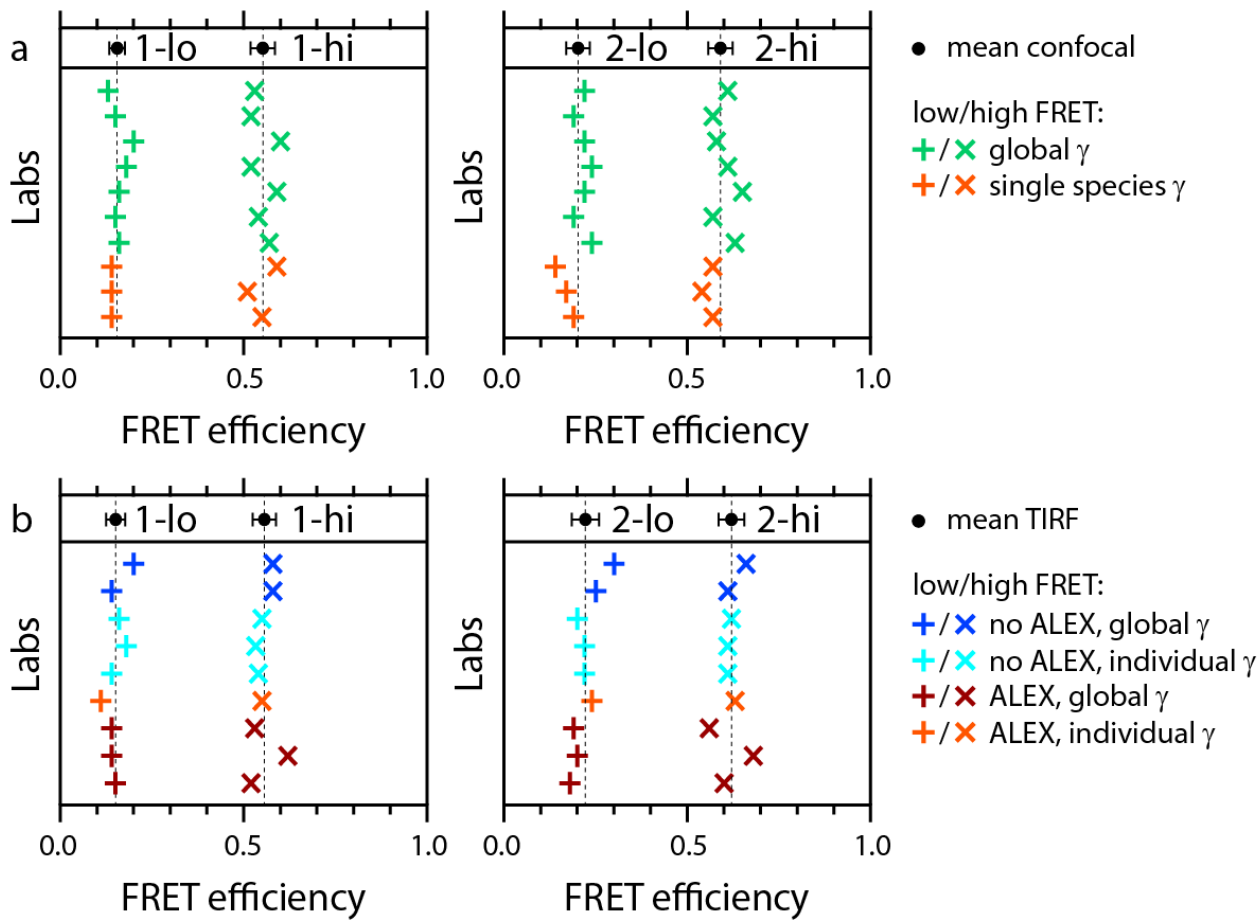

***Fig. 3:*** *Summary of the results of the intensity-based methods. **(a)** Confocal measurements. **(b)** TIRF measurements. Note that some laboratories performed measurement with both methods. In the top panel of each plot, the mean and standard deviations are depicted. Dashed lines indicate the means, their values are summarized in Table 4. Example correction factors are given in Table 2.*

**Distance determination**

The reported intensity-based FRET efficiencies are consistent throughout the labs, despite using different setup types and procedures. However, the ultimate goal is to derive distances from the FRET efficiencies. The efficiency-distance conversion requires both the knowledge of the Förster radius, $R_0$, for the specific FRET pair used, and a specific dye model, describing the behavior of the dye attached to the macromolecule. In the following, we describe (i) how $R_0$ can be determined and (ii) how to use a specific dye model to calculate the $R(\langle E \rangle)$ referred to as $R_{\langle E \rangle}$ and the $R_{MP}$. $R_{\langle E \rangle}$ is the apparent donor-acceptor distance, which is directly related to the experimental FRET efficiency $\langle E \rangle$ (eq. V), but it is not a physical distance. $R_{MP}$ is the distance between the mean positions of the dyes, which is not directly measured in the experiment, but is a real distance. $R_{MP}$ is important, for example, for mapping the physical distances required for rigid body structural modelling.

For computing $R_0$, the following parameters need to be determined or estimated (eq. VII): the index of refraction of the medium between the two fluorophores ($n$), the spectral overlap integral ($J$), the fluorescence quantum yield of the donor ($\Phi_{F,D}$), and the relative dipole orientation factor ($\kappa^2$) (see Online Methods for an estimate of their uncertainties). The overlap integral, the donor fluorescence quantum yield and the time-resolved fluorescence anisotropies of the donor-only and acceptor-only samples were measured by some of the participating labs (see *Table 3* and *Supplementary Table 1*). The dyes are attached via flexible linkers to the DNA, enabling rotation and translation within the accessible volume. The steady state anisotropies $r_s$ and residual anisotropies $r_\infty$ indicate that all used dyes are sufficiently mobile for dynamic averaging of the orientation factor, so that the average orientation factor $\langle \kappa^2 \rangle = 2/3$ can be assumed (the uncertainty in the exact average in $\kappa^2$ and the propagation of this error are further discussed in the *Online Methods Equation 22,23*). In short, this assumption is valid as long as the FRET rate ($k_{FRET}$) is much slower than the rotational relaxation rate ($k_{rot}$) of the dye. Our model assumptions also include that the dye with the translational diffusion rate ($k_{diff}$) samples the overall accessible volume within the experimental integration time ($1/k_{int}$), i.e. $k_{rot} \gg k_{FRET} \gg k_{diff} \gg k_{int}$. The validity of these assumptions is justified by experimental observables discussed in the *Online Methods, Section 3.4*.

The determined Förster radii for sample 1 and sample 2 are given in *Table 4*. Literature values differ mainly because the refractive index of water is often assumed, while we used $n_{im}$ = 1.40 here (see *Online Methods, Section 3.1*). Note that our careful error analysis led to an error estimate of 7% for the determined $R_0$, which is relatively large (mainly due to the uncertainty in $\kappa^2$). Thus, a rigorous error analysis for $R_0$ should be performed whenever distance information is needed (see the Online Methods).

Next, we used the measured smFRET efficiencies and the calculated Förster radii to compute the apparent distance $R_{\langle E \rangle}$ from each lab's data (eq. V) under the assumption $k_{rot} \gg k_{FRET} \gg k_{diff} \gg k_{int}$. *Figure 4a+b* shows the calculated values for these apparent distances for sample 1 and 2 for every data point in *Figure 3*. The average values for all labs are given in *Table 4*. Note that these errors only include the statistical variations of the FRET efficiencies, but do not include the error in the Förster radii, thus these errors represent the precision of the measurement, but not the accuracy. Including the knowledge of the dye attachment positions, a static structure of the DNA and this particular dye model, we computed also model values as described in *Supplementary Note 2*, which are also given in *Table 4*. Considering the error ranges, the experimental and model values agree very well with each other (the deviations range between 0 and 8 %).

The real distance between the center points (mean position) of the accessible volumes $R_{MP}$ deviates from the experimentally observed $R_{\langle E \rangle}$, because of the different averaging in distance and efficiency space. $R_{\langle E \rangle}$ corresponds to the real distance $R_{MP}$ only in the hypothetical case in which both dyes are unpolarized point sources, where the AVs have zero volume. In all other cases, $R_{MP}$ is the only physical distance. It can be calculated by one of the following two strategies: (i) If the dye model and the local environment of the dye is known (see *Figure 1*), simulation tools such as the FPS [8] can be used to compute the $R_{MP}$ from $R_{\langle E \rangle}$ for a given pair of AVs. (ii) If the structure of the investigated molecule is unknown *a priori*, a sphere is a useful assumption for the AV. A lookup table with polynomial conversion functions can then be calculated for defined AVs and Förster radii in order to relate $R_{\langle E \rangle}$ to $R_{MP}$ (*Supplementary Note 3*). The results, given as distances determined using the latter approach, are shown in *Figures 4c+d* and *Table 4*. The respective model values are based on the center points of the AVs depicted in *Figure 1* and given in *Table 4* (see *Supplementary Note 2* for details).

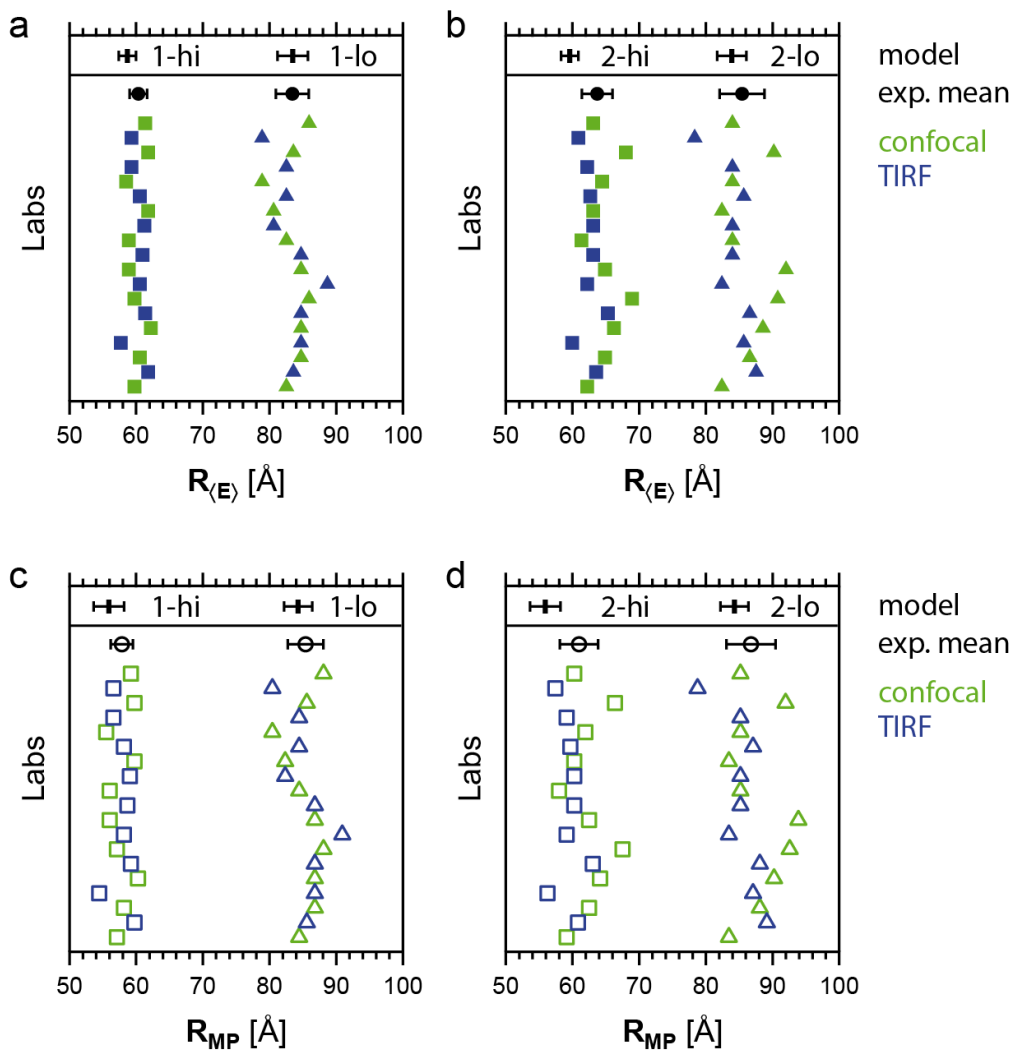

*Fig. 4: Mean inter-dye distances determined from the nineteen ⟨E⟩ values measured in sixteen different labs. (a) $R_{\langle E \rangle}$ for sample 1; (b) $R_{\langle E \rangle}$ for sample 2; (c) $R_{MP}$ for sample 1; (d) $R_{MP}$ for sample 2. The black dots (exp. mean) indicate the means and the error bars the statistical error (standard deviation) assuming $R_0=62.6$ Å and $R_0=68.0$ Å for sample 1 and 2, respectively). The black bars indicate the model values and their error (determined by variation of model parameters), see main text for details and Table 4 for values.*

**Distance uncertainties**

To understand the precision and accuracy of distance determination by smFRET, we estimated all uncertainty sources and propagated them into distance uncertainties. First, we discuss the error in determining the distance between two freely rotating but spatially fixed dipoles, $R_{DA}$, with the Förster equation (eq. III). *Figure 5a* shows how uncertainties in each of the correction factors ($\alpha, \gamma, \delta$) and the background signals ($BG_D, BG_A$) translate into the uncertainty of $R_{DA}$ (see *Supplementary Note 4* for all equations). The solid gray line shows the sum of these efficiency-dependent uncertainties, which are mainly setup-specific quantities. For the extremes of the distances the largest contribution to the uncertainty in $R_{DA}$ arises from background photons in the donor and acceptor channels. In the presented example with $R_0 = 62.6$ Å the total uncertainty $\Delta R_{DA}$ based on the *setup-specific uncertainties* is less than 4 Å for 35 Å < $R_{DA}$ < 90 Å. Notably, in confocal measurements, larger intensity thresholds can decrease this uncertainty. The uncertainty in $R_{DA}$ arising from errors in $R_0$ (blue line in *Figure 5b*) is added to the efficiency-related uncertainty in $R_{DA}$ (bold grey line) to estimate the total experimental uncertainty in $R_{DA}$ ($\Delta R_{DA,total}$, black line). The underlying uncertainties for determining $R_0$ are dominated by the dipole orientation factor $\kappa^2$ and the refractive index $n_{im}$, and are further discussed in the *Online Methods, Section 3.0*. Including the uncertainty in $R_0$, the error $\Delta R_{DA,total}$ for a single smFRET-based distance between two freely rotating point dipoles is less than 6 Å for 35 Å < $R_{DA}$ < 80 Å. It is worth mentioning that this uncertainty might be considerably reduced when multiple distances are calculated within a structure as the self-consistency in such a network enforces more precise localization [9]. In addition to the background issues, an $R_{DA}$ shorter than 30 Å may be prone to larger errors due to: (i) potential dye-dye interactions (ii) the dynamic averaging of the dipole orientations being reduced due to an increased FRET rate.

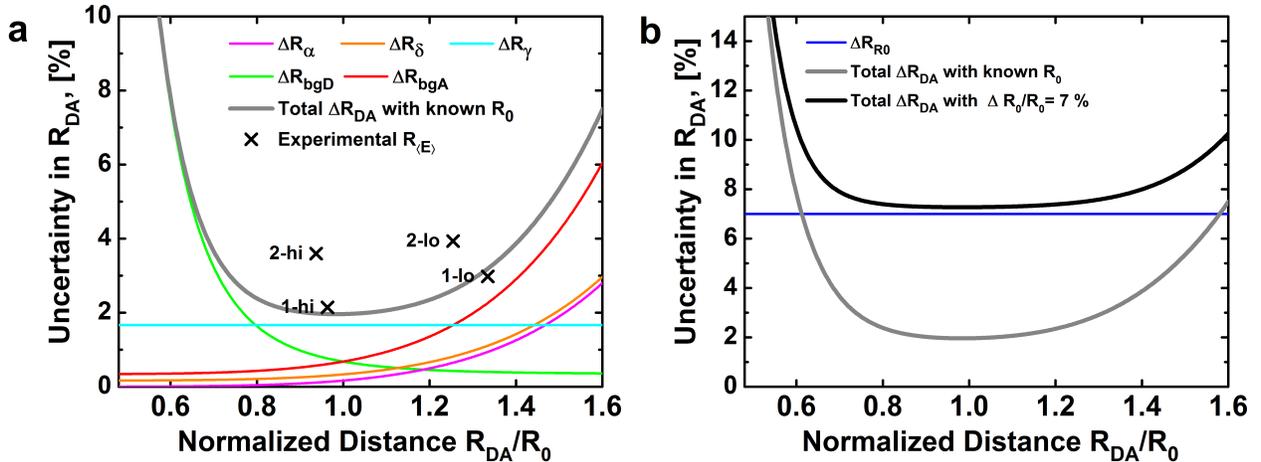

*Fig. 5:* Error propagation of experimental uncertainties. **(a)** $R_{DA}$ uncertainty contributions from the experimental correction factors: $\Delta R_\gamma$ (gamma factor), $\Delta R_{bgD}$ and $\Delta R_{bgA}$ (background), $\Delta R_\alpha$ (leakage), $\Delta R_\delta$ (direct excitation), total uncertainty with known $R_0$; crosses indicate uncertainty of experimental values of $R_{\langle E \rangle}$ across the labs. See Supplementary Note 4 for details on the error propagation. **(b)** Uncertainty in $R_{DA}$ (black line) based on the efficiency-related uncertainty (bold grey line) and the uncertainty for determining $R_0$ (blue line). See main text for details. Here we use the following uncertainties, which were determined for the confocal based measurements on sample 1: $\Delta R_0/R_0=7\%$, $\Delta\gamma/\gamma=10\%$, $\Delta I^{(BG)}/I=2\%$, $\Delta\alpha=0.01$ and $\Delta\delta=0.01$. For absolute values see Table 2.

It is important to verify the model assumption of a freely rotating and diffusing dye. Even though time-resolved fluorescence anisotropy can monitor whether dye rotation is fast, the possibility of dyes interacting with the DNA cannot be fully excluded. Thus, it is not clear if the dye molecule is completely free to sample the computed AV (free diffusion), or whether there are sites of attraction (preferred regions) or sites of repulsion (disallowed regions). In order to validate the dye model, we developed a self-consistency argument that bypasses several experimental uncertainties. The samples 1 and 2 were designed such that the ratio, $R_{rel}$, of their respective $R_{\langle E \rangle, lo}$ values is (quasi) independent of $R_0$:

$$R_{rel} = \frac{R_{\langle E \rangle, lo}}{R_{\langle E \rangle, hi}} = \frac{R_{0,lo}}{R_{0,hi}} \sqrt[6]{\frac{1/E_{lo} - 1}{1/E_{hi} - 1}} = \sqrt[6]{\frac{\kappa^2_{lo} \Phi_{D,lo} J_{lo} n^4_{hi}}{\kappa^2_{hi} \Phi_{D,hi} J_{hi} n^4_{lo}}} \sqrt[6]{\frac{1/E_{lo} - 1}{1/E_{hi} - 1}} = f \cdot \sqrt[6]{\frac{1/E_{lo} - 1}{1/E_{hi} - 1}}$$

The pre-factor, $f$, should be approximately the same for all measured dye combinations, for several reasons. First, the donor positions are identical for all lo- and hi-samples, respectively. Therefore, the following assumptions can be made: (i) the ratio of the donor quantum yields are identical; (ii) the ratio of the spectral overlaps $J$ for the lo- and hi- sample of one and the same dye pair should be the same; (iii) for the given geometry (see *Figure 1*) the refractive indices $n_{im}$ of the medium between the dyes should also be very similar; (iv) the ratio of the orientation factors $\kappa^2$ should be nearly equal as the measured donor anisotropies are low for the lo- and hi- positions. Second, the acceptor extinction coefficients eliminate each other as the acceptor is at the same position for the lo- and hi-samples. Thus, the different dye pairs and the model used in this study should all give similar values for $R_{rel}$. *Table 4* indeed shows that the relative $R_{\langle E \rangle}$ ratios are very similar and the deviations from the corresponding model values are less than 5% for sample 1 and 2, which is well within the experimental error. This further demonstrates the validity of the assumptions for the dye model and averaging regime used here.

While the analysis in this paper used a static model for the DNA structure, DNA is known not to be completely rigid [30]. We tested our DNA model by performing MD simulations of the DNA molecule (without attached dye molecules, see *Supplementary Note 6*) and found that the averaged expected FRET efficiency using the computed dynamically-varying DNA structure leads to comparable but slightly longer distances than for the static model. The deviations between the models and data are reduced (*Table 4*) for those cases where larger deviations were observed using static models.

**Discussion**

The aim of this blind study was to assess the consistency of FRET measurements from different labs around the world using various DNA FRET standards, without prior knowledge of the distance between the FRET pairs. The reported FRET efficiencies for the intensity-based measurements were consistent, with an overall standard deviation $\Delta E < \pm 0.05$ for each sample. This remarkable consistency was achieved by applying the same step-by-step procedure to perform the experiment and analyze the data.

We also showed that the factors required for correcting the FRET efficiency can be determined with high precision, independent of the different setup types and acquisition software used. Together the measurement errors in the correction factors cause an uncertainty in $R_{DA}$ of less than 5 %, which agrees well with the variations observed between the reported results from the different labs. Ultimately, we are interested in the absolute distances derived from these FRET measurements. *Figure 5b* shows good precision in the range from 0.6 $R_0$ to 1.6 $R_0$, which corresponds to an uncertainty of less than ±6 Å in the distance range from 35 to 80 Å for sample 1. This estimation is valid if the dyes are sufficiently mobile which has been supported by time-resolved anisotropy measurements and further confirmed by a self-consistency argument. For sample 2 the standard deviation is slightly larger than for sample 1 (see *Figure 5a*), which could be explained by dye specific photophysical properties. Interestingly for this case, the local gamma correction in the TIRF experiments yields data with a comparatively small standard deviation, while the global gamma correction data shows a comparatively large standard deviation (*Figure 3b*). The values for samples 3 and 4 (*Table 4*) show similar precision, considering the smaller number of measurements (*N*).

In addition to the achievable precision, we also tested the accuracy of the experimentally derived distances by comparing them with model distances. We found an excellent agreement within 2 Å for sample 1 and within 7 Å for sample 2. Both results are within the estimated theoretical distance uncertainty. For sample 2, which had the cyanine based dye Alexa647 instead of the carbopyronine based dye Atto647N as an acceptor, the lower accuracy could be explained by an imperfect sampling of the full AV or dye specific photophysical properties. Previously it has been shown that cyanine dyes are sensitive to their local environment [31] and therefore require especially careful characterization for each new labelled biomolecule.

The mean values of sample 3 and 4 are also within the error of the model values (for details see *Table 4* and *Supplementary Note 5*). This suggests that none of the four FRET pairs explored in this study exhibit significant dye artefacts. For future work, it will be very powerful to complement intensity-based high-precision smFRET studies with sm-lifetime studies because the picosecond time resolution can provide additional information on calibration and fast dynamic averaging.

The results from different labs and the successful self-consistency test clearly show the great potential of absolute smFRET-based distances for investigating biomolecular conformations and dynamics, as well as for integrative structural modelling. While $R_{MP}$ can be used to explore the real distance space in biomolecules, $R_{\langle E \rangle}$ can be used to directly compare the observables in FRET experiments with structural models. The ability to accurately determine distances on the molecular scale with smFRET experiments and to estimate the uncertainty of the measurements, provides the groundwork for smFRET-based structural and hybrid approaches. Together with the automated selection of the most informative pairwise labeling positions [16], and fast analysis procedures [8-10], we anticipate smFRET-based structural methods to become an important tool for *de novo* structural determination and structure validation, especially for large and flexible structures where other structural biology methods are difficult to apply.

## Acknowledgements

We thank the William Eaton Lab for early measurements that helped to design this study. We thank Thomas Peulen and Mykola Dimura and Prof. Rainer McDonald for stimulating discussions on FRET measurements, data analysis and modeling, and Bekir Bulat for measuring fluorescence quantum yields of Atto550 and 1-hi(Atto550). We also thank the company Atto-Tec for providing a reference sample of the dye Atto550 for the fluorescence characterization.

Parts of this work were funded by the European Research Council through ERC grant agreement no. 261227 to A.N.K., no. 671208 to C.A.M.S. and no. 681891 to T.H. J.M. acknowledges funding by the Deutsche Forschungsgemeinschaft (DFG) grant MI 749/4-1, P.T. by grant TI 329/10-1 and M.S. by grant SCHL 1896/3-1. B.S. acknowledges funding from the Swiss National Science Foundation. This work was supported by German Federal Ministry of Education and Research (BMBF) with 03Z2EN11 to M.S. J.H. acknowledges the Research Foundation Flanders (FWO) (grant G0B4915N). N.V. acknowledges the agency for Innovation by Science and Technology (IWT Flanders) for a doctoral scholarship. V.B. was supported by the Danish Council for Independent Research (Sapere Aude grant 0602-01670B). A.N.K. was supported by the UK BBSRC grant BB/H01795X/1. M.B. was supported by the National Institute of Mental Health grant MH081923. H.S., S.R.A. and I.S.Y. acknowledge support from start up funds from Clemson University. K.R.W. acknowledges NIH grants GM109832 and GM118508. B.W. gratefully acknowledges support by the Braunschweig International Graduate School of Metrology B-IGSM and the DFG Research Training Group GrK1952/1 "Metrology for Complex Nanosystems". T.D.C. was supported by start-up funds from the University of Sheffield. N.K.L. acknowledges the National Research Foundation of Korea funded by the Ministry of Science and ICT (NRF-2017R1A2B3010309).

## Author Contributions

B.H., T.H., J.M., C.S. designed the research; B.H., S.S., O.D., O.O., R.K., S.R.A., A.B., T.E., C.F., B.S., V.F., G.G., C.A.H., A.H., J.H., L.L.H., E.K., G.K., B.L., J.J.M., N.N., D.N., R.Q., C.R., B.S., H.S., L.S., S.T., N.V., B.W., I.S.Y. and T.D.C. performed measurements; B.H., S.S. and T.H. compared the measurements; all authors contributed to the analysis of the data and commented on the manuscript; B.H., S.S., T.D.C., J.M., C.S. and T.H. wrote the manuscript in consultation with O.D. and O.O.

## Competing financial interests

The authors declare no competing financial interests.

# Online Methods

*Table 1:* Nomenclature. Since the nomenclature for FRET-based experiments is not consistent, we propose and use the following terms in this manuscript.

| Central Definitions: | | |
|---|---|---|
| $E = \dfrac{F_{A\|D}}{F_{D\|D} + F_{A\|D}}$ | FRET efficiency | (I) |
| $S = \dfrac{F_{D\|D} + F_{A\|D}}{F_{D\|D} + F_{A\|D} + F_{A\|A}}$ | Stoichiometry | (II) |
| $E = \dfrac{1}{1 + R_{DA}^6/R_0^6}$ | FRET efficiency for a single donor acceptor distance $R_{DA}$ | (III) |
| $\langle E \rangle = \dfrac{1}{nm} \sum_{i=1}^{n} \sum_{j=1}^{m} \dfrac{1}{1 + \left\| \boldsymbol{R}_{A(j)} - \boldsymbol{R}_{D(i)} \right\|^6 / R_0^6}$ | Mean FRET efficiency for a discrete distribution of donor acceptor distances with the position vectors $\boldsymbol{R}_{D(i)}$ and $\boldsymbol{R}_{A(j)}$ | (IV) |
| $R_{\langle E \rangle} \equiv R(\langle E \rangle) = R_0(\langle E \rangle^{-1} - 1)^{1/6}$ | The apparent donor acceptor distance is computed from the average FRET efficiency for a distance distribution. It is a FRET averaged quantity which was also referred to as FRET-averaged distance $\langle R_{DA} \rangle_E$ (ref [32]). | (V) |
| $R_{MP} = \left\| \langle \boldsymbol{R}_{D(i)} \rangle - \langle \boldsymbol{R}_{A(j)} \rangle \right\|$ $= \left\| \dfrac{1}{n} \sum_{i=1}^{n} \boldsymbol{R}_{D(i)} - \dfrac{1}{m} \sum_{j=1}^{m} \boldsymbol{R}_{A(j)} \right\|$ | Distance between the mean dye positions with the position vectors $\langle \boldsymbol{R}_{D(i)} \rangle$ and $\langle \boldsymbol{R}_{A(j)} \rangle$ | (VI) |
| **Subscripts:** | | |
| $D$ or $A$ | Concerning donor or acceptor | |
| $A\|D$ | Acceptor fluorescence given donor excitation, $D\|D$, $A\|A$ accordingly | |
| $Aem\|Dex$ | Intensity in the acceptor channel given donor excitation, $Dem\|Dex$, $Aem\|Aex$, accordingly | |
| $app$ | apparent, i.e. including systematic, experimental offsets | |
| **Superscripts:** | | |
| $BG$ | Background | |
| $DO/AO$ | Donor-only species / Acceptor-only species | |
| $DA$ | FRET species | |
| $i$ -$iii$ | Indicates (i) the uncorrected intensity; (ii) intensity after BG correction; (iii) intensity after BG, alpha and delta corrections | |
| **Correction Factors:** | | |
| $\alpha = \dfrac{g_{R\|D}}{g_{G\|D}} = \dfrac{\langle {}^{ii}E_{app}^{(DO)} \rangle}{1 - \langle {}^{ii}E_{app}^{(DO)} \rangle}$ | Leakage of D fluorescence into A channel | |
| $\beta = \dfrac{\sigma_{A\|R}}{\sigma_{D\|G}} \dfrac{I_{Aex}}{I_{Dex}}$ | Normalization of excitation intensities, $I$, and cross-sections, $\sigma$, of A and D | |
| $\gamma = \dfrac{g_{R\|A}}{g_{G\|D}} \dfrac{\Phi_{F,A}}{\Phi_{F,D}}$ | Normalization of fluorescence quantum yields, $\Phi_F$, and detection efficiencies, $g$, of A and D | |
| $\delta = \dfrac{\sigma_{A\|G}}{\sigma_{A\|R}} \dfrac{I_{Dex}}{I_{Aex}} = \dfrac{\langle {}^{ii}S_{app}^{(AO)} \rangle}{1 - \langle {}^{ii}S_{app}^{(AO)} \rangle}$ | Direct acceptor excitation by the donor excitation laser (lower wavelength) | |

| | Primary Quantities: | |
|---|---|---|
| | $I$ | Experimentally observed intensity |
| $F$ | | Corrected fluorescence intensity |
| $\tau$ | | Fluorescence lifetime [ns] |
| $\Phi_{F,A}$ or $\Phi_{F,D}$ | | Fluorescence quantum yield of A and D, respectively |
| $r$ | | Fluorescence anisotropy |
| $R$ | | Inter-dye distance [Å] |
| | | Förster radius [Å], for a given J in units below (VII) |
| $\dfrac{R_0}{\text{Å}} = 0.2108 \sqrt[6]{\left(\dfrac{\Phi_{F,D}\kappa^2}{n_{im}^4}\right)\dfrac{J}{M^{-1}cm^{-1}nm^4}}$ | | |
| $\kappa^2 = (\cos\theta_{AD} - 3\cos\theta_D \cos\theta_A)^2$ | | Dipole orientation factor |
| $J = \int_0^\infty \bar{F}_D(\lambda)\,\varepsilon_A(\lambda)\lambda^4 d\lambda$ | | Spectral overlap integral [cm$^{-1}$M$^{-1}$nm$^4$] (see Supplementray Figure 6) |
| $\bar{F}_D(\lambda)$ with $\int_0^\infty \bar{F}_D(\lambda)\,d\lambda = 1$ | | Normalized spectral radiant intensity of the excited donor [nm$^{-1}$], defined as the derivative of the emission intensity F with respect to the wavelength. |
| $\varepsilon_A(\lambda)$ | | Extinction coefficient of A [M$^{-1}$ cm$^{-1}$] |
| $n_{im}$ | | Refractive index of the medium in-between the dyes |
| | Further Quantities | |
| $g_{R|A}$ or $g_{G|D}$ | | Detection efficiency of the red detector (R) if only acceptor was excited or green detector (G) if donor was excited. Analogous for others. |
| $\sigma_{A|G}$ | | Excitation cross-section for acceptor when excited with green laser. Analogous for the others. |
| | Abbreviations: | |
| | 1-lo | dsDNA oligo with Atto550 and Atto647N 23 basepairs apart |
| | 1-hi | dsDNA oligo with Atto550 and Atto647N 23 basepairs apart |
| | 2-lo | dsDNA oligo with Atto550 and Atto647N 15 basepairs apart |
| | 2-hi | dsDNA oligo with Atto550 and Alexa647 15 basepairs apart |

*Table 2:* Typical correction factors for sample 1 (Atto550-Atto647N) at given setups (reference lab). For the instrumental details of the setups see Supplementary Figures 3 and 4.

| Factor | Experiment type | |
|---|---|---|
| | confocal | TIRF |
| $\alpha$ | 0.11 | 0.07 |
| $\beta$ | 1.80 | 0.85 |
| $\gamma$ | 1.20 | 1.14 |
| $\delta$ | 0.11 | 0.065 |

*Table 3:* Typical parameters for sample 1 and sample 2 that define $R_0$ (Seidel lab). For their determination see Online Methods section 3.0.

| dye pairs | $\kappa^2$ | $n_{im}$ | $\Phi_{F,D}$ | $\varepsilon_A$ [M$^{-1}$cm$^{-1}$] | $J$ [cm$^{-1}$M$^{-1}$nm$^4$] | $R_0$ [Å] |
|---|---|---|---|---|---|---|
| Atto550-Atto647N | 2/3 | 1.40 | 0.765 | 150000 | $5.180\cdot 10^{15}$ | 62.6 |
| Atto550-Alexa647 | 2/3 | 1.40 | 0.765 | 270000 | $8.502\cdot 10^{15}$ | 68.0 |

**Table 4:** Summary of resulting mean efficiencies ⟨E⟩, apparent distance $R_{⟨E⟩}$, mean position distance $R_{MP}$ and corresponding model distances $R_{⟨E⟩}^{(model)}$ (Supplementary Note 2) and dynamic model distances $R_{⟨E⟩}^{(dynamic\ model)}$ (Supplementary Note 6) and the experimental ratio $R_{rel} = R_{⟨E⟩}^{(lo)}/R_{⟨E⟩}^{(hi)}$ and the model $R_{rel}^{(model)} = R_{⟨E⟩}^{(model,lo)} / R_{⟨E⟩}^{(model,hi)}$ for all intensity based measurements. The errors (standard deviations) report on the precision of the measurements and not their accuracy.

| Sample | N | ⟨E⟩ | $R_0$ [Å] | $R_{⟨E⟩}$ [Å] | $R_{⟨E⟩}^{(model)}$ [Å] | $R_{⟨E⟩}^{(dynamic\ model)}$ [Å] | $R_{rel}$ | $R_{rel}^{(model)}$ | $R_{MP}$ [Å] | $R_{MP}^{(model)}$ [Å] |
|---|---|---|---|---|---|---|---|---|---|---|
| 1-lo | 19 | 0.15±0.02 | 62.6±4.0 | 83.4±2.5 | 83.5±2.3 | 83.9 | 1.38 | 1.42 | 85.4±2.7 | 84.3±2.1 |
| 1-hi | 19 | 0.56±0.03 | | 60.3±1.3 | 58.7±1.3 | 60.3 | | | 57.9±1.7 | 55.9±2.3 |
| 2-lo | 19 | 0.21±0.04 | 68.0±5.0 | 85.4±3.4 | 83.9±2.2 | 84.2 | 1.34 | 1.41 | 86.8±3.7 | 84.3±2.1 |
| 2-hi | 19 | 0.60±0.05 | | 63.7±2.3 | 59.6±1.3 | 61.0 | | | 61.0±2.9 | 55.9±2.3 |
| 3-lo | 7 | 0.04±0.02 | 49.3[a] | 89.5±12.3 | 82.4±2.6 | 83.1 | 1.49 | 1.46 | 85.7±5.3 | 86.1±2.3 |
| 3-hi | 7 | 0.24±0.04 | | 60.1±2.3 | 56.4±1.9 | 58.4 | | | 61.1±2.9 | 57.1±2.3 |
| 4-lo | 4 | 0.13±0.06 | 57.0[a] | 79.6±6.2 | 82.6±2.5 | 83.5 | 1.31 | 1.43 | 82.9±6.8 | 86.3±2.3 |
| 4-hi | 4 | 0.41±0.04 | | 60.7±1.7 | 57.6±1.5 | 59.5 | | | 60.4±2.3 | 58.7±2.3 |

[a] The $R_0$ for these samples have been taken from the literature and converted from a refractive index of $n_{im}$ = 1.33 to $n_{im}$ = 1.40:
Sample 3: $R_0$ = 49.3 Å from ref. [33]
Sample 4: $R_0$ = 57.0 Å from ref. [24]

## 1. Samples

Altogether, 8 different FRET-samples were designed with the acceptor dyes positioned 15 or 23 base pairs away from the donor dyes (see *Table 5* and *Supplementary Note 5*). IBA GmbH (Göttingen) synthesized and labeled the single DNA strands followed by HPLC purification. Here the dyes were attached to a thymidine (dT), which is known to cause the least fluorescence quenching of all nucleotides[26].

Most labs measured the following four DNA samples listed in *Table 5*. Therefore, we focus on these four samples in the main text of this manuscript. The additional samples and the corresponding measurements can be found in *Supplementary Note 5* and *Supplementary Figure 2* and *Table 4*. The following buffer was requested for all measurements: 20 mM MgCl$_2$, 5 mM NaCl, 5 mM Tris, at pH 7.5, degassing just before the measurement at room temperature.

The linker lengths were chosen in a way that all dyes had about the same number of flexible bonds between the dipole axis and the DNA. The Atto550, Alexa647 and Atto647N already have an intrinsic flexible part before the C-linker starts (*Supplementary Figure 1*). In addition, the DNAs were designed such that the distance ratio between the high FRET efficiency and low FRET efficiency sample should be the same for all samples, largely independent of $R_0$.

*Table 5.* The main focus in the manuscript are the 1-lo, 1-hi, 2-lo, 2-hi samples. The so called donor strand (D-strand) is labeled with donor dye and acceptor strand (A-strand) with acceptor dye. The labeling sites of the donor and acceptor are shown in green and in red on the sequence respectively.

| Name | Base position (Linker), strand | Dyes (Donor/Acceptor) | Sequence |
|---|---|---|---|
| **1-lo** | T 31(C2), D-strand <br> T 31(C2), A-strand | Atto550 NHS Ester/ <br> Atto647N NHS | 5'– GAG CTG AAA GTG TCG AGT TTG TTT GAG TGT **T**TG TCT GG – 3' <br> 3'– CTC GAC T**T**T CAC AGC TCA AAC AAA CTC ACA AAC AGA CC – 5'-biotin |
| **1-hi** | T 23(C2), D-strand <br> T 31(C2), A-strand | Atto550 NHS Ester/ <br> Atto647N NHS | 5'– GAG CTG AAA GTG TCG AGT TTG T**T**T GAG TGT TTG TCT GG – 3' <br> 3'– CTC GAC T**T**T CAC AGC TCA AAC AAA CTC ACA AAC AGA CC – 5'-biotin |
| **2-lo:** | T 31(C2), D-strand <br> T 31(C2), A-strand | Atto550 NHS Ester/ <br> Alexa647 NHS Ester | 5'– GAG CTG AAA GTG TCG AGT TTG TTT GAG TGT **T**TG TCT GG – 3' <br> 3'– CTC GAC T**T**T CAC AGC TCA AAC AAA CTC ACA AAC AGA CC – 5'-biotin |
| **2-hi:** | T 23(C2), D-strand <br> T 31(C2), A-strand | Atto550 NHS Ester/ <br> Alexa647 NHS Ester | 5'– GAG CTG AAA GTG TCG AGT TTG T**T**T GAG TGT TTG TCT GG – 3' <br> 3'– CTC GAC T**T**T CAC AGC TCA AAC AAA CTC ACA AAC AGA CC – 5'-biotin |

## **2. General correction procedure**

Efficiency *E* and Stoichiometry *S* are defined in *Table 1*. Determination of the corrected FRET *E* and *S* is largely based on the Lee et al. approach [21] and consists of the following steps: (1) data acquisition; (2) generation of uncorrected *E* vs *S* 2D histograms; (3) background subtraction; (4) correction for the position-specific excitation in TIRF experiment; (5) correction for leakage and direct acceptor excitation; (6) correction for excitation intensities and absorption cross-sections, quantum yields and detection efficiencies.

**2.1. Data acquisition:** The sample with both dyes is measured and the three intensity time traces are extracted: acceptor emission upon donor excitation ($I_{Aem|Dex}$), donor-emission upon donor excitation ($I_{Dem|Dex}$), and acceptor-emission upon acceptor excitation ($I_{Aem|Aex}$).

*For the confocal* setups a straight forward burst identification is performed by binning the trace into 1 ms bins. Usually a minimum threshold (e.g. 50 photons) is applied to the sum of the donor and acceptor signals upon donor excitation for each bin. This threshold is used again in every step, such that the number of utilized bursts may change from step to step (if the γ correction factor is not equal to one). Some labs use sophisticated burst-search algorithms. For example, the dual channel burst search[34,35] recognizes the potential bleaching of each dye within bursts. Note that the choice of the burst-search algorithm can have an influence on the γ correction factor. For standard applications, the simple binning method is often sufficient, especially for well-characterized dyes and low laser powers. This study shows that the results do not significantly depend on these conditions (if applied properly), as every lab used its own setup and procedure at this stage.

*For the TIRF* setups, traces with one acceptor and one donor are selected, defined by a bleaching step. In addition, only the relevant *range* of each trajectory – i.e. prior to photo-bleaching of either dye - is included in all further steps.

**2.2. 2D histogram:** A 2D histogram (*Figure 2a,b*) of the apparent stoichiometry, $^iS_{app}$, vs. apparent FRET efficiency, $^iE_{app}$, defined by Equations 1 and 2 is generated, where

$$^iS_{app} = (I_{Aem|Dex} + I_{Dem|Dex})/(I_{Aem|Dex} + I_{Dem|Dex} + I_{Aem|Aex}) \quad (1)$$

$$^iE_{app} = I_{Aem|Dex}/(I_{Aem|Dex} + I_{Dem|Dex}) \quad (2)$$

**2.3. Background correction:** Background $I^{(BG)}$ is removed from each uncorrected intensity $^iI$ separately, leading to the background corrected intensities $^{ii}I$, $^{ii}S_{app}$, $^{ii}E_{app}$:

$$\begin{aligned}
^{ii}I_{Dem|Dex} &= {}^iI_{Dem|Dex} - I^{(BG)}_{Dem|Dex} \\
^{ii}I_{Aem|Aex} &= {}^iI_{Aem|Aex} - I^{(BG)}_{Aem|Aex} \\
^{ii}I_{Aem|Dex} &= {}^iI_{Aem|Dex} - I^{(BG)}_{Aem|Dex}
\end{aligned} \quad (3)$$

*For confocal measurements*, the background is determined by averaging the photon count rate for all time bins that are below a threshold, which is e.g. defined by the maximum in the frequency vs intensity plot (density of bursts should not be too high). Note, that a previous measurement of only the buffer can uncover potential fluorescent contaminants, but it can differ significantly from the background of the actual measurement. The background intensity is then subtracted from the intensity of each burst in each channel (eq. 3). Typical background values are 0.5-1 photon / ms (*Figure 2a*).

*For TIRF measurements*, various trace-wise or global background corrections can be applied. The most common method defines background as the individual offset (time average) after photo-bleaching of both dyes in each trace. Another possibility is to select the darkest spots in the illuminated area and to subtract an average background time trace from the data or to use a local background, e.g. with a mask around the particle. The latter two have the advantage that possible (exponential) background bleaching is also corrected for. We have not investigated the influence of the kind of background correction during this study, but a recent study has shown that not all background estimators are suitable for samples with a high molecule surface coverage [36].

Overall, a correction of the background is very important, but can be done very well in different ways.

**2.4. The position specific excitation correction (optional for TIRF):** The concurrent excitation profiles of both lasers are key for accurate measurements (see *Supplementary Figure 5*). Experimental variations across the field of view are accounted for using a position-specific normalization:

$$_{(profile)}^{ii}I_{Aem|Aex} = {}^{ii}I_{Aem|Aex} \frac{I_D(x',y')}{I_A(x,y)} \quad (4)$$

where $I_D(x',y')$ and $I_A(x,y)$ denote the excitation intensities at corresponding positions in the donor or acceptor image, respectively. Individual excitation profiles are determined as the mean image of a stack of images recorded while moving across a sample chamber with dense dye coverage.

**2.5. Leakage (α) and direct excitation (δ):** After the background correction, the leakage fraction of the donor emission into the acceptor detection channel and the fraction of the direct excitation of the acceptor by the donor-excitation laser are determined. The correction factor for leakage (α) is determined by Equation 5 using the FRET efficiency of the donor-only population (D-only in *Figure 2a,b(ii)*). The correction factor for direct excitation (δ) is determined by Equation 6 from the stoichiometry of the acceptor-only population (A-only in *Figure 2a,b(ii)*).

$$\alpha = \frac{\langle {}^{ii}E_{app}^{(DO)}\rangle}{1-\langle {}^{ii}E_{app}^{(DO)}\rangle} \tag{5}$$

$$\delta = \frac{\langle {}^{ii}S_{app}^{(AO)}\rangle}{1-\langle {}^{ii}S_{app}^{(AO)}\rangle} \tag{6}$$

where ${}^{ii}E_{app}^{(DO)}$ and ${}^{ii}S_{app}^{(AO)}$ are calculated from the background-corrected intensities ${}^{ii}I$ of the corresponding population, i.e. donor-only or acceptor-only, respectively. This correction together with the previous background correction results in the donor-only population being located at $\langle E\rangle = 0, \langle S\rangle = 1$ and acceptor-only population at $\langle S\rangle = 0, \langle E\rangle = 0 \ldots 1$. The corrected acceptor fluorescence after donor excitation $F_{A|D}$ is given by Equation 7, which yields the updated expressions for the FRET efficiency and stoichiometry, Equations 8 and 9, respectively.

$$F_{A|D} = {}^{ii}I_{Aem|Dex} - \alpha \, {}^{ii}I_{Dem|Dex} - \delta \, {}^{ii}I_{Aem|Aex} \tag{7}$$

$${}^{iii}E_{app} = F_{A|D}/(F_{A|D} + {}^{ii}I_{Dem|Dex}) \tag{8}$$

$${}^{iii}S_{app} = (F_{A|D} + {}^{ii}I_{Dem|Dex})/(F_{A|D} + {}^{ii}I_{Dem|Dex} + {}^{ii}I_{Aem|Aex}) \tag{9}$$

In principle, the leaked donor signal could be added back to the donor emission channel[37]. However, this requires precise knowledge about spectral detection efficiencies, which is not otherwise required, and has no effect on the final accuracy of the measurement. As the determination of $\alpha$ and $\delta$ influences the $\gamma$ and $\beta$ correction in the next step, both correction steps can be repeated in an iterative manner if required (e.g. if the $\gamma$ and $\beta$ factors deviate largely from one).

**2.6. $\gamma$ and $\beta$ correction factors:** Differences in the excitation intensities and cross-section, as well as, quantum yields and detection efficiencies are accounted for by using the correction factors $\gamma$ and $\beta$, respectively. If the fluorescence quantum yields do not depend on efficiencies or such dependence is negligible (*homogenous approximation*), mean values of efficiencies $\langle {}^{iii}E_{app}^{(DA)}\rangle$ and of stoichiometries $\langle {}^{iii}S_{app}^{(DA)}\rangle$ are related by equation 10:

$$\langle {}^{iii}S_{app}^{(DA)}\rangle = \left(1 + \gamma\beta + (1-\gamma)\beta\,\langle {}^{iii}E_{app}^{(DA)}\rangle\right)^{-1} \tag{10}$$

So, in the homogeneous approximation, $\gamma$ and $\beta$ correction factors can be determined by fitting FRET populations to the ${}^{iii}S_{app}^{(DA)}$ vs. ${}^{iii}E_{app}^{(DA)}$ histogram with the line defined by Equation 10. As this method relies on the analysis of ${}^{iii}S_{app}^{(DA)}$, ${}^{iii}E_{app}^{(DA)}$ values obtained from multiple species, we term this method, *global $\gamma$-correction*. Such a fit can be performed for all FRET populations together, for any of their subsets, and in principle, for each single-species population separately (see below). Alternatively, a linear fit of inverse $\langle {}^{iii}S_{app}^{(DA)}\rangle$ vs. $\langle {}^{iii}E_{app}^{(DA)}\rangle$ with y-intercept $a$ and slope $b$ can be performed.

In this case, $\beta = a + b - 1$ and $\gamma = (a-1)/(a+b-1)$.

Error propagation, however, is more straightforward if Equation 10 is used. If there is a complex dependence between properties of dyes and efficiencies, the homogeneous approximation is no longer applicable. In this case, the relationship between ${}^{iii}S_{app}^{(DA)}$, ${}^{iii}E_{app}^{(DA)}$ for different populations (or even subpopulations for the same single-species) cannot be described by Equation 10 with a single $\gamma$ correction factor. Here, $\beta$ and $\gamma$ can be determined for a single species. We call this *"single-species $\gamma$-correction"*. This works only if the efficiency broadening is dominated by distance fluctuations. The reason for this assumption is the dependency of these corrections factors on both the stoichiometry and the distance-dependent efficiency. In our study, global and local $\gamma$-correction yielded similar results. Therefore, the homogenous approximation and distance fluctuations as the main cause for efficiency broadening can be assumed for sample 1 and 2. Systematic variation of the $\gamma$-correction factor yields an error of about 10%.

Alternatively, determination of $\gamma$, $\beta$ factors can be done trace-wise, e.g. as in msALEX experiments[38] where the $\gamma$ factor is determined as the ratio of the decrease in acceptor signal and the increase in donor signal upon acceptor bleaching. We call such an alternative correction, *individual $\gamma$-correction*. The analysis of local distributions can provide valuable insights about properties of the studied system.

After $\gamma$ and $\beta$ correction, the corrected donor (acceptor) fluorescence after donor (acceptor) excitation $F_{D|D}$ ($F_{A|A}$) amounts to:

$$F_{D|D} = \gamma \, {}^{ii}I_{Dem|Dex} \qquad (11)$$

$$F_{A|A} = \frac{1}{\beta} \, {}^{ii}I_{Aem|Aex} \qquad (12)$$

**2.7. Fully corrected values:** Application of all corrections leads to the estimates of real FRET efficiencies, $E$, and stoichiometries, $S$, from the background corrected intensities, ${}^{ii}I$. The explicit expressions of fully corrected FRET efficiency and stoichiometry are:

$$E = \frac{\left[{}^{ii}I_{Aem|Dex} - \alpha \, {}^{ii}I_{Dem|Dex} - \delta \, {}^{ii}I_{Aem|Aex}\right]}{\gamma\left[{}^{ii}I_{Dem|Dex}\right] + \left[{}^{ii}I_{Aem|Dex} - \alpha \, {}^{ii}I_{Dem|Dex} - \delta \, {}^{ii}I_{Aem|Aex}\right]} \qquad (13)$$

$$S = \frac{\gamma\left[{}^{ii}I_{Dem|Dex}\right] + \left[{}^{ii}I_{Aem|Dex} - \alpha \, {}^{ii}I_{Dem|Dex} - \delta \, {}^{ii}I_{Aem|Aex}\right]}{\gamma\left[{}^{ii}I_{Dem|Dex}\right] + \left[{}^{ii}I_{Aem|Dex} - \alpha \, {}^{ii}I_{Dem|Dex} - \delta \, {}^{ii}I_{Aem|Aex}\right] + 1/\beta\left[{}^{ii}I_{Aem|Aex}\right]} \qquad (14)$$

Plots of the $E$ vs. $S$ histogram are shown in *Figure 2a(iv)* and *2b(iv)*. Now, the FRET population should be symmetric to the $S = 0.5$ line. The donor-only population should still be located at $E = 0$ and the acceptor-only population at $S = 0$. Finally, the corrected FRET efficiency histogram is generated using events with a stoichiometry of $0.3 < S < 0.7$ (see *Figure 2a,b* bottom). The expected value of the corrected FRET efficiencies $\langle E \rangle$ is deduced as the center of a Gaussian fit to the efficiency histogram. This is a good approximation for FRET efficiencies in the range between about 0.1 and 0.9. In theory, the shot-noise limited efficiencies follow a binomial distribution if the photon number per burst is constant. For extreme efficiencies or data with a small average number of photons per burst, the efficiency distribution can no longer be approximated with a Gaussian. In this case and also in the case of efficiency broadening due to distance fluctuations, a detailed analysis of the photon statistics can be useful [34,39-41].

### 3. Uncertainty in distance due to $R_0$:

According to Förster theory[1], the FRET efficiency, $E$, and the distance, $R$, are related by equation III in *Table 1*. In this study, we focused on comparing $E$ in a blind study across different labs. The Seidel lab determined an $R_0$ for this system to convert efficiencies to distances. There are many excellent reviews on how to determine the Förster radius $R_0$ [20,42,43] and a complete discussion would be beyond the scope of this experimental comparison study. In the following, we estimate and discuss the different sources of uncertainty in $R_0$, by utilizing standard error propagation (see also *Supplementary Note 4*). $R_0$ is given by equation VII, *Table 1*.

The 6th power of the Förster radius is proportional to the relative dipole orientation factor $\kappa^2$, the donor quantum yield $\Phi_{F,D}$, the overlap integral $J$, as well as $n^{-4}$, where $n$ is the refractive index of the medium:

$$R_0^6 \sim \kappa^2 \cdot \Phi_{F,D} \cdot J \cdot n^{-4} \qquad (15)$$

For Figure 5b, we use a total Förster radius related distance uncertainty of 7 %, which is justified by the following estimate. Please note that the error in the dipole orientation factor is always very specific for the investigated system, while the errors in the donor quantum yield, overlap integral and refractive index are more general, but their mean values do also depend on the environment.

**3.1. The refractive index.** Different values for the refractive index in FRET systems have been used historically, but ideally the refractive index of the donor-acceptor intervening medium $n_{im}$ should be used, though some experimental studies suggest that the use of the refractive index of the solvent may be appropriate, but this is still open for discussion (see e. g. discussion in[44]).

$$R_0^6(n) \sim n_{im}^{-4} \qquad (16)$$

In the worst case, this value $n_{im}$ might be anywhere in-between the refractive index of the solvent ($n_{water}$ = 1.33) and a refractive index for the dissolved molecule ($n < n_{oil}$=1.52) [45], i.e. $n_{water} < n_{im} < n_{oil}$. This would result in a maximum uncertainty of $\Delta n_{im} < 0.085$. As recommend by Clegg, we used $n_{im}$ = 1.40 to minimize this uncertainty[46] (see *Table 3*). The distance uncertainty propagated from the uncertainty of the refractive indices can then be assumed to be:

$$\Delta R_0(n) \approx \frac{4}{6} R_o \frac{\Delta n_{im}}{n} < 0.04 \cdot R_0 \qquad (17)$$

**3.2. The donor quantum yield** $\Phi_{F,D}$ is position dependent, therefore we measured the fluoresence lifetimes and quantum yields of the free dye Atto550 and the 1-hi and 1-lo(Atto550) (see *Supplementary Table 1*).

In agreement with Sindbert et al. [32], the uncertainty of the quantum yield is estimated at $\Delta\Phi'_{F,}$ = 5 % arising from the uncertainties of the $\Phi_F$ values reference dyes and the precision of the absorption and fluorescence measurements. Thus, the distance uncertainty owing to the quantum yield is estimated at:

$$\Delta R_0(\Phi_{F,D}) \approx \frac{R_0}{6} \frac{\Delta \Phi_{F,D}}{\Phi_{F,D}} = 0.01 \cdot R_0 \qquad (18)$$

**3.3. The overlap integral $J$** was measured for the unbound dyes in solution (Atto550 and Atto647N), as well as for samples 1-lo and 1-hi. This resulted in a deviation of about 10 % for $J$ using the literature values for the extinction coefficients. All single stranded labeled DNA samples used in this study were purified with HPLC columns providing a labelling efficiency of at least 95 %. The label efficiencies of the single stranded singly-labeled DNA and of the double stranded singly-labeled DNA samples were determined by the ratio of the absorption maxima of the dye and the DNA and were all above 97 %. This indicates an error of the assumed exctinction coefficient of less than 3 %. Thus, the distance uncertainty due to the overlap spectra and a correct absolute acceptor extinction coefficient can be estimated by Equation 19. However, the uncertainty in the acceptor extinction coefficient might be larger for other environments, such as when bound to a protein.

$$\Delta R_0(J) \approx \frac{R_0}{6} \frac{\Delta J}{J} = 0.025 \cdot R_0 \qquad (19)$$

In addition to the above uncertainty estimation, the $J$-related uncertainty can also be obtained by verifiing the self-concistency of a $\beta$-factor network [9]. Finally, we found little uncertainty by using the well tested dye Atto647N. Fluorescence spectra were measured on a Fluoromax4 spectrafluorimeter (Horiba, Germany). Absorbance spectra were recorded on a Cary5000 UV/VIS spectrometer (Agilent, USA). See *Supplementary Figure 6*.

**3.4. The $\kappa^2$ factor and model assumptions:** The uncertainty in the distance depends on the dye model used [15]. Several factors need to be considered, given the model assumptions of unrestricted dye rotation, equal sampling of the entire accessible volume, and the following rate inequality $k_{rot} >> k_{FRET} >> k_{diff} >> k_{int}$.

First, the use of $\langle \kappa^2 \rangle = 2/3$ is justified if $k_{rot} >> k_{FRET}$, because then there is rotational averaging of the dipole orientation during energy transfer. $k_{rot}$ is determined from the rotational correlation time $\rho_1 < 1$ ns and $k_{FRET}$ is determined from the fluorescence lifetimes 1 ns $< \tau_{fl} < 5$ ns. Hence the condition $k_{rot} >> k_{FRET}$ is not strictly fulfilled. We estimate the error this introduces into $\kappa^2$ from the time-resolved anisotropies of donor and acceptor dyes. If the transfer rate is smaller than the fast component of the anisotropy decay (rotational correlation time) of donor and acceptor. Then, the combined anisotropy, $r_C$, is given by the residual donor and acceptor anisotropies ($r_{D,\infty}$ and $r_{A,\infty}$, respectively):

$$r_C = \sqrt{r_{A,\infty}} \sqrt{r_{D,\infty}} \tag{20}$$

In theory, the donor and the acceptor anisotropy should be determined at the time of energy transfer. If the transfer rate is much slower than the fast component of the anisotropy decay of donor and acceptor, the residual anisotropy can be used (*Supplementary Fig. 8*)[9]. Also, the steady state anisotropy values can give an indication of the rotational freedom of the dyes on the relevant time scales, if the inherent effect by the fluorescence lifetimes is taken into account (see Perrin equation, *Supplementary Table 1* and *Supplementary Figure 7*).

If the steady-state anisotropy and $r_C$ are low ($< 0.2$), one can assume (but not prove) sufficient isotropic coupling (rotational averaging), i.e. $\langle \kappa^2 \rangle = 2/3$, with an uncertainty of about 5 %[9]:

$$\Delta R_0(\kappa^2, r_C < 0.2) \approx 0.05 \cdot R_0 \tag{21}$$

**3.5. Spatial sampling.** In addition, it is assumed that both dyes remain in a fixed location for the duration of the donor lifetime, i.e. $k_{FRET} >> k_{diff}$: where $k_{diff}$ is defined as the inverse of the diffusion time through the complete AV. Recently the diffusion coefficient for a tethered Alexa488 dye was determined to be D=10 Å$^2$/ns (ref. [26]). Therefore, $k_{diff}$ is smaller than the $k_{FRET}$. For short distances ($< 5$ Å) the rates become comparable, but the effect on the inter dye distance distribution within the donor lifetime is small, as has been observed in time resolved experiments. We also assumed that, in the experiment, the efficiencies are averaged for all possible inter-dye positions. This is the case when $k_{diff} >> k_{int}$, which is a very good assumption for TIRF experiment with $k_{int} > 100$ ms and also for confocal experiments with $k_{int}$ around 1 ms.

**3.6. Overall uncertainty in $R_0$.** Time-resolved anisotropy measurements of samples 1 and 2 resulted in combined anisotropies below 0.1. Therefore we assumed isotropic coupling to obtain $R_{MP}$. The $R_{MP}$ match the model distances very well, further supporting these assumptions. Finally, an experimental study on $\kappa^2$ distributions also obtained typical errors of 5 %[32].

The overall uncertainty for the Förster radius would then result in:

$$\Delta R_0(n^{-4}, \Phi_{F,D}, J, \kappa^2) = \sqrt{\Delta R_0(n)^2 + \Delta R_0(\Phi_{F,D})^2 + \Delta R_0(J)^2 + \Delta R_0(\kappa^2)^2} \lesssim 0.07 \cdot R_0 \tag{22}$$

The absolute values determined for this study are summarized in *Table 3*. Please note that the photophysical properties of dyes vary in different buffers and when attached to different biomolecules. Therefore, all four quantities contributing to the uncertainty in $R_0$ should be measured for the system under investigation. When supplier values or values from other studies are applied, the uncertainty can be much larger. The values specified here could be further evaluated and tested in another blind study.